\documentclass[%
reprint,
amsmath,amssymb,prl,
aps, longbibliography, superscriptaddress
]{revtex4-1}

\usepackage{graphicx}
\usepackage{dcolumn}
\usepackage{bm}
\usepackage{braket}
\usepackage{mathtools}
\usepackage{bbm}
\usepackage{comment}
\usepackage{makecell}
\usepackage{afterpage}
\usepackage{amsthm}
\usepackage{siunitx}
\usepackage{times}
\usepackage{cellspace}

\usepackage{hyperref}
\hypersetup{
	colorlinks=true,
	linkcolor=blue,
	citecolor=blue,
	filecolor=magenta,
	urlcolor=cyan
}

\def\iu{{\rm i}}

\def\dif{{\rm d}}
\def\pdif{{\mathcal{D}}}

\usepackage[normalem]{ulem}

\graphicspath{{.}{..}}

\begin{document}
	\title{Signatures of fractional statistics in nonlinear pump-probe spectroscopy}
	\date{\today}
	\def\oxf{{Rudolf Peierls Centre for Theoretical Physics, Clarendon Laboratory, Parks Road, Oxford OX1 3PU, United Kingdom}}
	
	\author{Max McGinley}
	\affiliation{\oxf}
	\affiliation{T.C.M. Group, Cavendish Laboratory, JJ Thomson Avenue, Cambridge CB3 0HE, United Kingdom}
	\author{Michele Fava}
	\affiliation{\oxf}
	\affiliation{Philippe Meyer Institute, Physics Department, \'Ecole Normale Sup\'erieure (ENS),
		Universit\'e PSL, 24 rue Lhomond, F-75231 Paris, France}
	\author{S.A. Parameswaran}
	\affiliation{\oxf}
	
	\begin{abstract}
		We show that the presence of anyons in the excitation spectrum of a two-dimensional system can be inferred from nonlinear spectroscopic quantities. In particular, we consider pump-probe spectroscopy, where a sample is irradiated by two light pulses with an adjustable time delay between them. The relevant response coefficient exhibits a universal form that originates from the statistical phase acquired when anyons created by the first pulse braid around those created by the second. This behaviour is shown to be qualitatively unchanged by non-universal physics including non-statistical interactions and small nonzero temperatures. In magnetic systems, the signal of interest can be measured using currently available terahertz-domain probes, highlighting the potential usefulness of nonlinear spectroscopic techniques in the search for quantum spin liquids.
	\end{abstract}
	
	\maketitle
	
	Two dimensional quantum systems can host emergent excitations that are `anyonic', falling outside the usual classification of bosonic vs.~fermionic statistics \cite{Leinaas1977, Wilczek1982}. Their theoretical discovery has triggered a search for physical systems that can support such excitations, motivated jointly by their potential relevance in understanding highly correlated quantum materials such as the cuprate high-temperature superconductors \cite{Lee2007}, and the possibility of using such systems to perform fault-tolerant quantum computation \cite{Nayak2008}. In solid-state settings, the primary foci of these investigations are 
	semiconductor heterostructures in the fractional quantum Hall regime \cite{Arovas1984, Stormer1999} and quantum spin liquids (QSLs) in frustrated magnetic systems \cite{Knolle2019,Broholm2020}.
	
	
	Vital to this search is the identification of experimental signatures of anyons and of the topologically ordered phases of matter that underpin them. One approach is to look for excitations that carry quantum numbers smaller than those of the underlying microscopic degrees of freedom (e.g.~charges smaller than $e$), which have to be created in groups of more than one at a time. This `fractionalization' of excitations, which is often concomitant with anyonic statistics, has been 
	established in semiconductor heterostructures using shot noise measurements \cite{Saminadayar1997,Picciotto1998}; similarly, a broad continuum in the dynamical spin structure factor, as measured through inelastic neutron scattering, serves as evidence favouring fractionalization in certain candidate QSLs \cite{Han2012,Banerjee2017}.
	
	However, the direct detection of fractional statistics, as opposed to just fractionalization,
	necessitates a setup where braiding of excitations actually occurs, with the statistical phases being detected by interferometric means \cite{Chamon1997, Safi2001, Kim2005, Halperin2011, Campagnano2012, Rosenow2016}. This has been achieved in quantum Hall systems only recently \cite{Bartolomei2020,Nakamura2020}, using novel geometries to guide anyons along edge modes. Such an approach to braiding is not feasible in QSLs due to limitations in the sample geometries that can be fabricated,  and so appropriate bulk signatures must be found instead. Scattering of neutrons and light (either electron spin resonance or Raman scattering) are the most commonly employed methods to probe the dynamics of electron spins (or pseudospins). However, since 
	most theoretical studies have focused on linear response quantities \cite{Cepas2008,Qi2009,Punk2014,Knolle2014, Knolle2014a, Kamfor2014, Knolle2015, Zschocke2015,Morampudi2020}---primarily the dynamical spin structure factor, which only witnesses
	fractionalization---it is not yet clear how non-trivial braiding of anyons can be inferred from such techniques.
	
\begin{figure}
		\centering
		\includegraphics{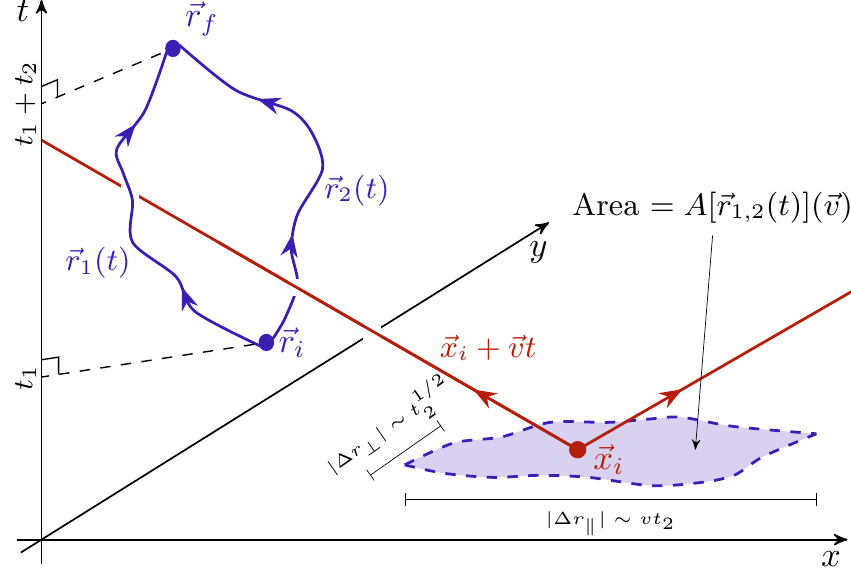}
		\caption{The area functional $A[\vec{r}_{1,2}(t)](\vec{v})$ measures the range of initial coordinates $\vec{x}_i$ for which the pump trajectory $\vec{x}(t) = \vec{x}_i + \vec{v} t$ links with the spacetime loop formed by $\vec{r}_{1,2}(t)$---only these trajectories contribute to $\chi_{\rm PP}$ [Eq.~\eqref{eq:ChiDef}]. At large times, this area scales as $t_2^{3/2}$, leading to Eq.~\eqref{eq:main-result}.}
		\label{fig:traj}
	\end{figure}

	In this work, we unveil how the braiding statistics of excitations in QSLs manifest themselves in \textit{nonlinear} spectroscopic quantities. In particular, we study pump-probe spectroscopy, where the system is perturbed by a sequence of two pulses with a tunable time delay between them. In a QSL, where excitations are deconfined, anyons created by the first pulse can move and braid with those created by the second (Fig.~\ref{fig:traj}). We show that the statistical phase induced by this process leads to a universal relationship between the linear response coefficient $\chi^{(1)}(t_2)$ and the pump-probe signal $\chi_{\rm PP}^{(3)}(t_1, t_2)$ (a nonlinear response function which we define later in the paper):
	\begin{align}
		\chi_{\rm PP}^{(3)}(t_1, t_2) = \chi^{(1)}(t_2) \times c_{\rm PP}\Big[t_2^{3/2} + o(t_2^{3/2})\Big],
		\label{eq:main-result}
	\end{align}
	where $c_{\rm PP}$ is a non-universal constant. Eq.~\eqref{eq:main-result} serves as a fingerprint of anyonic excitations. The relevant signals can be directly measured using currently available terahertz-domain spectroscopic techniques, which have previously been used to probe ultrafast magnetization dynamics in systems with spontaneous spin ordering \cite{Yamaguchi2010, Kampfrath2011, Mukai2016, Lu2017}. Remarkably, even though such probes only couple to zero-momentum operators, i.e.~no spatial resolution is available, the physics of braiding statistics can still be seen.
	
	Despite the strongly interacting nature of the systems under consideration, our key result---the universal relationship \eqref{eq:main-result}---can be understood intuitively using a semiclassical argument, akin to those given for one-dimensional systems in Refs.~\cite{Sachdev1997,Fava2022}. In short, we relate the behaviour of the pump-probe signal to the probability that anyons braid, which itself grows asymptotically as $t_2^{3/2}$, where $t_2$ is the time between creation and annihilation of the probe anyons; this leads to Eq.~\eqref{eq:main-result}.
	
	
	In this short paper, we introduce the relevant quantities and systems to be studied, and present the intuitive arguments outlined above. These are backed up by more rigorous, concrete calculations in a companion paper \cite{LongPaper}, wherein we also discuss the implications of our results on the nature of thermal relaxation in topologically ordered systems. The findings reported here are shown to be qualitatively unchanged when additional factors are accounted for, such as non-statistical interactions between quasiparticles and nonzero temperatures.
	
	Given the robust, universal nature of our result and the fact that the relevant signal can be measured using pre-existing experimental techniques, we propose that pump-probe spectroscopy can serve as a helpful diagnostic of quantum spin liquid physics, complementing those based on conventional linear spectroscopy.


	\textit{Linear and pump-probe spectroscopy.---}
	{
		In this paper we will be concerned with a particular form of nonlinear response known as pump-probe spectroscopy \cite{Yan1989, Mukamel1995}. First, let us briefly review conventional light-based linear spectroscopy. Here, a sample in an equilibrium state $\hat{\rho}_0$ is perturbed by a weak pulse, which we will dub ``probe''. Subsequently, the light re-emitted from the sample is measured, the signal from which can be used to infer the dynamical correlator
		$\chi^{(1)}(t_1,t_2) = L^{-2}\braket{\hat{A}_2(t_1+t_2) \hat{A}_1(t_1)}_0$ (see e.g. Ref.~\cite{Mukamel1995}).
		Here $\hat{A}_{1,2}(t) = e^{\iu \hat{H} t} \hat{A}_{1,2} e^{-\iu \hat{H} t}$ are time-evolved operators in the Heisenberg picture, the expectation value $\langle\, \cdot \, \rangle_0 $ is with respect to $\hat{\rho}_0$, and we have divided by the system volume $L^2$ to ensure that $\chi^{(1)}$ is intensive.
		The operators $\hat{A}_{1,2}$ depend on the specific light-matter coupling in the system, and our findings do not depend on their details. However, as customary in solid-state systems, we assume that $\hat{A}_{1,2}$ only contains zero-wavevector ($q = 0$) components.
	}

	In pump-probe spectroscopy, the system is first brought out of equilibrium by an initial `pump' pulse of light, before the probe pulse is applied.
	This initial pulse is typically short and intense, and so we can describe its effect as an instantaneous unitary rotation of the initial equilibrium state, which without loss of generality can be assumed to occur at time $t = 0$. Explicitly,
	\begin{align}
		\rho_0 \xrightarrow[\text{pump pulse}]{t=0} \hat{\rho}_{\rm pert} \coloneqq e^{-\iu \kappa \hat{A}_0} \rho_0 e^{\iu \kappa \hat{A}_0}
		\label{eq:Pump}
	\end{align}
	where $\hat{A}_0$ is the operator to which the pump pulse couples, and $\kappa$ is a constant controlling the strength of the pulse. The experiment is repeated with and without the pump pulse, and the pump-probe signal is defined as the difference between the dynamical correlation function in each case, again normalized by the system volume $L^2$
	\begin{align}
		\chi_{\rm PP}(t_1, t_2) &= \frac{1}{L^2}\braket{ \hat{A}_2(t_1+t_2)\hat{A}_1(t_1)}_{\rm pert} \hspace*{-2pt} - \chi^{(1)}(t_1,t_2). 
		\label{eq:ChiDef}
	\end{align}
	Here, the expectation value $\braket{\,\cdot\,}_{\rm pert}$ is evaluated using the perturbed state \eqref{eq:Pump}.  
	
	As a final remark on these spectroscopic techniques, we note that the sequence of pulses described above has already been implemented in experiments investigating ultrafast magnetization dynamics of materials with spontaneous spin ordering \cite{Yamaguchi2010, Kampfrath2011, Mukai2016, Lu2017}, enabled by recent developments in the generation of high-intensity THz-domain pulses with short time resolution \cite{Blanchard2007,Yeh2007}. Having defined and motivated the study of $\chi_{\rm PP}$, we now move on to investigating its behaviour in two-dimensional topologically ordered systems.

	\textit{Low-energy effective theory.---}Eq.~\eqref{eq:ChiDef} can be applied quite generally, provided that one knows the system Hamiltonian $\hat{H}$ and the operators $\hat{A}_{0,1,2}$ to which the electromagnetic fields couple. These of course depend on the specifics of the system in question. However, since we are interested in gapped topologically ordered 2D systems, the dynamics at sufficiently small temperatures and frequency ranges---equivalently, large times $t_{1,2}$---is well-described by an effective low-energy theory of stable quasiparticle excitations. Thus, rather than using a particular microscopic description of $\hat{A}_{0,1,2}$ and $\hat{H}$, we will compute $\chi_{\rm PP}$ using this effective quasiparticle theory, the structure of which is constrained by the topological phase in which the system is in. 
	
	While our results are applicable to any two-dimensional topological phase, for concreteness and clarity we will make reference to the $\mathbbm{Z}_2$ spin liquid phase, which is the phase to which the toric code belongs \cite{Kitaev1997, Kitaev2003}. (The generalization to other phases, including non-Abelian phases, can be found in our companion paper.) Such systems possess two flavours of quasiparticle excitations, termed electric ($e$) and magnetic ($m$) particles. Each particle is bosonic with respect to itself, but electric and magnetic excitations are mutual semions, meaning that a phase $e^{\iu \pi} = -1$ is acquired whenever one moves in a loop around the other. 
	
	The effects of the operators $\hat{A}_{0,1,2}$ will be to create or annihilate groups of quasiparticles located within a few correlation lengths $\xi$ of each other. Due to their topological nature, individual electric or magnetic quasiparticles cannot be created by local operators in isolation; instead, they must be formed in pairs $(e,e)$ or $(m,m)$. We can therefore approximate the action of $\hat{A}_i$ as
	\begin{align}
		\hat{A}_i \ket{\text{VAC}} = \sum_{\alpha = e, m} a_{i,\alpha} \int \dif^2 \vec{r} \ket{\vec{r}, \vec{r}\,}_{\alpha, \alpha}+\dots,
		\label{eq:QuasiparticleCreation}
	\end{align}
	where $\ket{\vec{r}, \vec{r}\,'}_{\alpha, \beta}$ is a two-quasiparticle state with an excitation of flavour $\alpha$ ($\beta$) at position $\vec{r}$ ($\vec{r'}$), and $a_{i,\alpha}$ are coefficients that determine the weights of electric- and magnetic-anyon pairs, which we treat as phenomenological parameters. Here we used that $\hat{A}_i$ is a zero-momentum operator, i.e. it acts uniformly over all space. Terms with more than two anyons, which we denote by ``\dots'', give subleading corrections at long times.

	Once created, these quasiparticles will generically be mobile, having some dispersion relation $\epsilon_\alpha(\vec{k})$ determined by the Hamiltonian $\hat{H}$, where $\alpha = e, m$ labels the excitation flavours. For simplicity we will assume $\epsilon(\vec{k}) = \Delta_\alpha + |\vec{k}|^2/2m_\alpha + O(k^3)$, although the only important aspect is that the Hessian of $\epsilon(\vec{k})$ is non-zero at $\vec{k}=0$.
	
	Having described the structure of the $\mathbbm{Z}_2$ spin liquid quasiparticle theory and the coupling operators $\hat{A}_{0,1,2}$, we are ready to compute the relevant response coefficients, starting with the simplest case: linear response. For the time being, we will set the temperature $T = 0$, such that $\hat{\rho}_0 = \ket{\text{VAC}}\bra{\text{VAC}}$, and neglect interactions between quasiparticles, besides the statistical interactions that are mediated through braiding phases. These assumptions will be lifted later.
	
	\textit{Linear response from $\mathbbm{Z}_2$ quasiparticle theory.---}The linear response coefficient $\chi^{(1)}(t_1, t_2) = L^{-2}\braket{\hat{A}_2(t_1+t_2) \hat{A}_1(t_1)}_0$ can be reduced to a two-quasiparticle problem using Eq.~\eqref{eq:QuasiparticleCreation}. Starting from the vacuum state, the operator $\hat{A}_1$ creates a pair of quasiparticles at some spacetime location $(\vec{r}_1, t_1)$, which then propagate and are later annihilated together by the operator $\hat{A}_2$ at $(\vec{r}_2, t_2)$; here the coordinates $\vec{r}_{1,2}$ are to be integrated over according to \eqref{eq:QuasiparticleCreation}. Invoking spatial- and time-translational invariance, for each flavour of quasiparticle $\alpha$ we have a contribution $\chi^{(1)} \propto  \int \dif^2\vec{r} \braket{\vec{r}, \vec{r}|e^{-\iu \hat{H} t_2}|\vec{0}, \vec{0}}_{\alpha, \alpha}$. At long times $t_2$, this expression only depends on the quasiparticle dispersion near the band minimum, and so after making the substitution $\epsilon_\alpha(k) \approx \Delta_\alpha + \vec{k}^2/2m_\alpha$ we find $\chi^{(1)}(t_1, t_2) \propto e^{-2\iu \Delta_\alpha t_2}\times t_2^{-1}$. This power law decay of the linear response coefficient is due to the decreasing amplitude of finding both quasiparticles at the same point in space, which is necessary for their annihilation due to Eq.~\eqref{eq:QuasiparticleCreation}.

	Although the above computation of the linear response function is straightforward, when it comes to calculating the analogous pump-probe coefficient it will be more convenient to adopt a path integral formulation of the quasiparticle dynamics. Formally,  we can write
	\begin{align}
		\chi^{(1)}(t_1, t_2) = \int \dif^2 \vec{r}_f \int_{r_{1,2}(t_1)=0}^{r_{1,2}(t_1+t_2)=r_f}\hspace*{-5pt} \pdif \vec{r}_1(t) \pdif \vec{r}_2(t) e^{\iu S_0[\vec{r}_{1,2}(t)]}
		\label{eq:LinPathInteg}
	\end{align}
	where $S_0[\vec{r}_1(t), \vec{r}_2(t)] = (m_\alpha/2)\int_{t_1}^{t_1+t_2} \dif t(|\dot{r}_1(t)|^2 + |\dot{r}_2(t)|^2)$ is the real-time Feynman action for the probe quasiparticles with a quadratic dispersion, and the path integral is taken over all trajectories $\vec{r}_{1,2}(t)$ of the excitations, subject to the constraints on the initial and final positions. 
	
	\textit{Pump-probe response coefficient.---}The effect of the pump pulse is to create a population of additional quasiparticles at time $t = 0$, which we refer to as `pump anyons' to distinguish them from the `probe anyons' created by the probe pulse at $t=t_1$. While a non-perturbative approach can be employed (see Ref.~\cite{LongPaper}), here for simplicity we will expand to leading order in the pump pulse intensity $\chi_{\rm PP}(t_1, t_2) = \kappa^2 \chi^{(3)}_{\rm PP}(t_1, t_2) + O(\kappa^3)$ (following standard notation in nonlinear response theory). When we expand the post-pump state \eqref{eq:Pump} to order $\kappa^2$, the only terms that survive involve the creation of a single pair of pump anyons. If this pair has the same flavour as the probe anyons, i.e.~both electric or both magnetic, then the presence of these additional quasiparticles will have no bearing on the subsequent dynamics of the probe anyons, and the first term in Eq.~\eqref{eq:ChiDef} will exactly cancel with the second. We therefore consider terms where the pump anyons are electric ($e$), and the probe anyons are magnetic ($m$); the same considerations apply \textit{vice-versa}. These two species of quasiparticle are mutual semions, so a phase of $e^{\iu \pi}$ is incurred whenever pump and probe anyons braid. 
	
	Once created, the motion of the pump anyons can be described semiclassically \cite{Sachdev1997, Fava2022} (this approximation is justified in our companion paper). This means the anyons are modelled as wavepackets of negligible spatial extent moving at a constant speed set by their group velocity $\vec{v}_k = \partial_k \epsilon_e(\vec{k})$. The velocities of a pair will be opposite, $\pm \vec{v}$, and so the trajectories of these excitations will be $\vec{x}(t) = \vec{x}_i \pm \vec{v} t$. The initial coordinate $\vec{x}_i$ is uniformly distributed over all space by virtue of $\hat{A}_0$ being translation invariant [Eq.~\eqref{eq:QuasiparticleCreation}],
	while the velocity $\vec{v}$ will be distributed in a way that depends on the microscopic structure of $\hat{A}_0$ and the dispersion $\epsilon_e(\vec{k})$. The form of this distribution $p(\vec{v})\dif^2 \vec{v}$ will not be important here, and so we leave it unspecified.

	For large enough $t_1$, the pump anyons will be well-separated by the time the probe anyons are created; thus, processes where two pump anyons braid with the probe anyons can be ignored. Therefore, let us focus on the trajectory of a single pump anyon, ignoring the other, and consider the probability that a braid indeed occurs. This requires values of $\vec{x}_i$ and $\vec{v}$ for which the pump anyon trajectory threads the spacetime loop formed by the probe anyon trajectories $\vec{r}_{1,2}(t)$; see Fig.~\ref{fig:traj}. For a fixed $\vec{r}_{1,2}(t)$, we denote the probability of this occurring as $P[\vec{r}_{1,2}(t)]$.
	Then, up to a multiplicative factor, $\chi_{\rm PP}^{(3)}$ is given by the same path integral as in Eq.~\eqref{eq:LinPathInteg}, but with an additional weighting of $(-2)\times P[\vec{r}_{1,2}(t)]$. Here, the factor of $-2 \equiv e^{\iu\pi}-1$ is due to the subtraction of the two terms in \eqref{eq:ChiDef}, the latter of which is the same as the first but with all braiding phases set to unity.

	The contribution to $P[\vec{r}_{1,2}(t)]$ coming from pump anyons with a particular velocity $\vec{v}$ is proportional to the area spanned by coordinates $\vec{x}_i$ such that the line $\vec{x}_i + \vec{v} t$ passes through the spacetime loop formed by $\vec{r}_{1,2}(t)$. This area, which is the blue shaded region in Fig.~\ref{fig:traj}, is a functional of $\vec{r}_{1,2}(t)$ and a function of $\vec{v}$; we denote it $A[\vec{r}_{1,2}(t)](\vec{v})$. (Interestingly, if one sets $\vec{v} = \vec{0}$ then one obtains a functional that appeared in the study of a particle in a random magnetic field \cite{Altshuler1992}.) Note that the resulting expression is $t_1$-independent, since, since shifts of $t_1$ are equivalent to rigid translations of the pump anyon trajectories $\vec{x}(t)$. Altogether, we have $P[\vec{r}_{1,2}(t)] \propto \int \dif^2\vec{v}\,p(\vec{v}) A[\vec{r}_{1,2}(t)](\vec{v})$.

	Using this expression for $P[\vec{r}_{1,2}(t)]$, the pump-probe response function can actually be evaluated exactly via a lengthy calculation---see our companion paper for details. However, we can quickly obtain its asymptotic behaviour by estimating the typical size of $A[\vec{r}_{1,2}(t)](\vec{v})$ for those trajectories $\vec{r}_{1,2}(t)$ that contribute the most to the path integral \eqref{eq:LinPathInteg}. Defining the component of $\vec{x}_i$ parallel (perpendicular) to $\vec{v}$ as $x_\parallel$ ($x_\perp$), we wish to determine a typical range over which these coordinates can be varied while ensuring the trajectories still link. Firstly, $x_\perp$ will span a range of the order of $|r_{1,\perp}(t) - r_{2,\perp}(t)|$ for some intermediate time $t_1 < t < (t_1+t_2)$, where $r_{1,\perp}(t)$ is the component of $\vec{r}_1(t)$ perpendicular to $\vec{v}$. By dimensional analysis of the Feynman path integral, this distance can be seen to scale asymptotically as $\sim \sqrt{t_2/m}$ (see also Ref.~\cite{Altshuler1992}). As for the parallel component, a shift $x_\parallel \rightarrow x_\parallel + a$ is equivalent to a time-translation of the pump anyon trajectory $t \rightarrow t + a/v$. Hence, $x_\parallel$ will be varied over a range $\sim |v|t_2$. Combining these, we find $A[\vec{r}_{1,2}(t)](\vec{v}) \sim |\vec{v}| t_2^{3/2}$ in the limit of large $t_2$. When we average over the distribution of velocities $ p(\vec{v})\dif^2\vec{v}$, this result gives us $P[\vec{r}_{1,2}(t)] \sim t_2^{3/2}$. Evidently, as $t_2$ increases, the probability of a braid event goes up, due to the increased spacetime area spanned by the trajectories $\vec{r}_{1,2}$.
	
	As mentioned above, $\chi^{(3)}_{\rm PP}(t_1, t_2)$ only differs from $\chi^{(1)}(t_2)$ through the presence of a factor of $P[\vec{r}_{1,2}(t)]$ in the path integral. When we replace $P[\vec{r}_{1,2}(t)]$ with its typical value as estimated above, we obtain our main result, Eq.~\eqref{eq:main-result}. (A more detailed calculation in our companion paper confirms that this scaling behaviour does indeed arise when the full path integral is evaluated.)


	
	In this specific case, if we combine Eq.~\eqref{eq:main-result} with the linear response coefficient derived earlier, we find
	\begin{align}
		|\chi_{\rm PP}(t_1, t_2)| \propto \underbrace{\frac{1}{t_2}}_{\text{recombination}} \times \underbrace{t_2^{3/2}}_{\text{probability of braiding}} = t_2^{1/2}.
		\label{eq:SqrtT}
	\end{align}
	Interestingly, the probability of braiding increases at a faster rate than the decay of the pump-probe response coefficient due to anyons diffusing apart. This means that the pump-probe response actually grows with time, and will eventually reach a size where perturbation theory breaks down \cite{LongPaper}. 
	
	\textit{Discussion.---}Our result \eqref{eq:main-result} has been derived based on general considerations about the creation and propagation of quasiparticle excitations, along with the fusion rules and braiding relations of the $\mathbbm{Z}_2$ spin liquid topological phase. We propose that this is a generic feature of all two-dimensional topologically ordered phases, both Abelian and non-Abelian, and that such behaviour continues to hold in the presence of nonuniversal short-ranged interactions and/or small nonzero temperatures. Concrete justifications of this statement are given in our companion paper \cite{LongPaper}; here we summarise some intuitive arguments that support our claim.
	
	In other topologically ordered phases, the operators $\hat{A}_{0,1,2}$ will create multiplets of anyons whose mutual braiding phases may be different. If the braiding relationship between pump and probe anyons is a phase different from $e^{\iu \pi}$ (but still nontrivial), then all that changes is the prefactor of $(e^{\iu \pi}-1)$, and \eqref{eq:main-result} still holds. Moreover, since the probe anyons become well-separated for large $t_1$, the key physics is not affected if the pump anyons are not mutually bosonic.
	
	
	If there are nontrivial braiding (or exchange) phases between probe anyons, then our representation of the linear response coefficient [Eq.~\eqref{eq:LinPathInteg}] will have to be modified to account for the statistical interactions between anyons~\cite{Morampudi2017}. 
	Although the these differences will change the behaviour of $\chi^{(1)}$, we emphasise that the path integral representation of the pump-probe response coefficient will be modified in exactly the same way. Importantly, the path integral representations of the two response coefficients still only differ by an additional factor proportional to $P[\vec{r}_{1,2}(t)]$. Therefore, provided that typical values of $P[\vec{r}_{1,2}(t)]$ scale in the same way with $t_2$ (which can be argued on dimensional grounds \cite{LongPaper}), we will find that \eqref{eq:main-result} still holds. 
	
	
	We now consider the impact of non-statistical interactions and finite temperatures, which were heretofore neglected. As for the former, interactions within the pump anyons have no bearing once pairs have separated at large enough $t_1$, as was the case for intra-pump-anyon braiding phases; similarly, interactions between probe anyons alter $\chi^{(1)}$ and $\chi_{\rm PP}$ in exactly the same way, which as we saw above does not change the asymptotic scaling \eqref{eq:main-result}. We are then left with interactions between pump and probe multiplets. The probability that a probe anyon is deflected by a pump anyon can be estimated using scattering theory---if $\vec{v}$ is the pump anyon velocity then $P_{\rm scat} \sim |\vec{v}|t_2 \sigma \lambda$, where $\lambda$ is the area density of anyons created by the pump pulse and $\sigma$ is the scattering cross-section \cite{Fava2022}.  This scales slower than the $t_2^{3/2}$ scaling of the braiding probability, and so the contributions to $\chi_{\rm PP}$ coming from non-statistical interactions are subleading compared to the dominant signal \eqref{eq:main-result}. The contrasting behaviour the braiding vs.~scattering probabilities is due to the fact that scattering requires excitations to be within some small distance of each other, whereas braiding can occur even when all particles remain far apart from each other at all times.
	
	
	At finite temperatures, there will be an excess population of anyons present in both the linear and pump-probe response functions. These can braid with and/or scatter the probe anyons, which generally has the effect of suppressing the two-point correlation function. As it turns out, this suppression leads to modulation of the response coefficients by a `squished exponential' $\exp(-[t_2/\tau_T]^{3/2})$, where $\tau_T$ is a non-universal temperature-dependent timescale---see our companion paper for details. Importantly, once again the linear and pump-probe response coefficients are affected in exactly the same way, and so Eq.~\eqref{eq:main-result} is unaffected; however to experimentally confirm this one will need to find a time window such that the asymptotic form has been reached, but that the decay of each signal due to thermally activated anyons is not so strong such that their magnitudes are immeasurably small. 
	
	\textit{Experimental implications.---}Our results demonstrate that fractional statistics can be inferred from time domain measurements, even without spatial resolution. This establishes the feasibility of a new class of future experiments for the detection of QSLs, which offer more conclusive signatures than inferences based on the diffusivity of inelastic neutron scattering cross-sections.
	The question of whether a definitive enough signature can be seen in any given experiment will depend on the specifics of the material being investigated and the temporal resolution of the chosen spectroscopic methods; thus to assess near-term applications, it is instructive to focus on a particular QSL candidate currently under investigation. We consider $\alpha$-$\text{RuCl}_3$, where there is evidence for a magnetic field-induced gapped QSL phase \cite{Yadav2016, Sears2017, Baek2017, Kasahara2018} (the anyons in this putative phase are non-Abelian, but the above arguments still hold \cite{LongPaper}). 
	
	Observing the universal signature \eqref{eq:main-result} requires the existence of a window of times $t_2$ where transient contributions [$o(t_2^{3/2})$] have subsided, but attenuation of the signal due to extraneous effects is minimal. Transients arise from anharmonicities in the dispersion $\epsilon_\alpha(k)$, which can be neglected for times beyond the inverse bandwidth of the anyons $(\Delta \epsilon)^{-1}$. Based on estimates of the magnetic interaction strength $J \sim 70$-$\SI{90}{\kelvin}$ for $\alpha$-$\text{RuCl}_3$ \cite{Banerjee2017}, along with DFT bandstructure calculations \cite{Hou2017}, this requires $t_2 \gtrsim \SI{1}{\pico\second}$. Assuming that non-statistical anyon-anyon interactions are short-ranged, impurity scattering is the only mechanism by which the signal can decay at strictly zero temperature. Using an estimate of the scattering rate $\tau_{\rm scat}^{-1} = J(a/\lambda_{\rm mf})$, where $\lambda_{\rm mf}$ is the mean free path, we see that such effects only set in after a much longer time $t_2 \gg J^{-1}$, provided that $\lambda_{\rm mf}$ is much greater than the lattice spacing $a$, which is indeed satisfied for any reasonable disorder strength. At nonzero temperatures, scattering between thermal and pump/probe anyons leads to decay of the signal, but using an estimate of the excitation gap obtained from neutron scattering data $\Delta \approx \SI{2}{\milli\electronvolt}$ \cite{Banerjee2018}, this should be highly suppressed at operating temperatures $T < \SI{20}{\kelvin}$. Given that picosecond-resolved nonlinear spectroscopy measurements have already been achieved in magnetic materials \cite{Yamaguchi2010, Kampfrath2011, Mukai2016, Lu2017}, we anticipate that the signal \eqref{eq:main-result} should be detectable using temporal resolutions and temperature regimes that are currently accessible.
	
	As a final remark, we note that due to the $q = 0$ nature of the signal, the universal physics should be present irrespective of the sample geometry, and possibly even in powdered samples.

	\begin{acknowledgements}
		\textit{Acknowledgements.---}We thank Nick Bultinck, Claudio Castelnovo, John Chalker, Rahul Nandkishore, and Steven H. Simon for useful discussions, and Sarang Gopalakrishnan, Romain Vasseur, and Fabian Essler for discussions and collaboration on related work. We acknowledge support from the European Research Council under the European Union Horizon 2020 Research and Innovation Programme, Grant Agreement No. 804213-TMCS, and from the UK Engineering and Physical Sciences Research Council via  grant EP/S020527/1. Statement of compliance with EPSRC policy framework on research data: This publication is theoretical work that does not require supporting research data.
	\end{acknowledgements}

	\bibliography{nonlinear_anyon}
	
\end{document}